\documentstyle[preprint,aps]{revtex}	 

\begin{document}
\draft

\title{Proportion Regulation in Globally Coupled Nonlinear Systems}
\author{
Tsuyoshi Mizuguchi
\footnote{e-mail address: gutchi@sawada.riec.tohoku.ac.jp}
and
Masaki Sano
\footnote{e-mail adderss: sano@sawada.riec.tohoku.ac.jp}
}
\address{Research Institute of Electrical Communication \\
Tohoku University, Sendai 980-77, Japan}

\maketitle

\begin{abstract}
As a model of proportion regulation in
differentiation process of biological system,
globally coupled activator-inhibitor systems are studied.
Formation and destabilization of
one and two cluster state are predicted analytically.
Numerical simulations show that the proportion of units of clusters
is chosen within a finite range and
it is selected depend on the initial condition.
\end{abstract}

\pacs{87.10.+e,87.22.-q,87.22.As}


The regulation of proportion among different cell types in a tissue is
a general and important aspect of biological development.
It is well known that the proportion between the two different cell types
is roughly constant irrespective of the slug size of cellular slime mold
{\it Dictyostelium discoideum} (Dd) amoebae
\cite{Raper,Bonner,Loomis,Takeuchi}.
Initially the same type of aggregative cells, when dissociated,
randomly mixed, and reaggregated, differentiate into two types
cells (prespore and prestalk cells) without pattern formation.
It is known now that cell differentiation starts independently on the
cell position, and later cell sorting forms the two-zoned prestalk-prespore
pattern in slug of Dd
\cite{Loomis,Takeuchi,Ozaki}.
Similar regulation mechanism can be observed in
caste populations of social insects such as ants and bees
\cite{Frisch,Wilson}.
In the division of work, proportion is regulated irrespective of the size
of society nor the artificial partial extinction by the experimenter.

No theoretical model exists to describe the proportion regulation.
Any pattern formation model such as Turing type instability with diffusive
coupling \cite{Turing} is incompatible with the observation that
the Dd cells start to differentiate independent to their positions.
A large population of identical units interacting equally to the other
units (globally coupled nonlinear system) is a good candidate to
describe the phenomena.
It is an idealized model of the cases, when the diffusion length of
chemical factor, e.g. differentiation inducing factor (DIF) or
pheromone, is large enough compared to the cell size, or when
the individual units moves around to interact with others.

Recently, globally coupled chaotic map\cite{Kaneko,Wiesenfeld}
and globally coupled oscillators
\cite{Kuramoto,Golomb,Hakim,Okuda,Nakagawa_Kuramoto,Sakaguchi}
are studied and interesting phenomena including clustering and their
destabilization are observed.
However, analysis of cluster state is
difficult for these systems because the unit itself is complex enough.
It is also doubtful that chaos or oscillation is playing essential roles
in proportion regulation of biological system such as Dd.
In this respect, a minimum model of clustering is preferable.

Our model of globally coupled system is composed of $N$ activator-inhibitor
type units which have two variables $u$ and $v$.
The dynamics of each unit is modeled as
\begin{eqnarray}
&&\begin{array}{rcccl}
\dot u_j & = & au_j - bv_j & - u_j^3 & +K_1(\overline u -u_j),\\
\dot v_j & = & cu_j - dv_j &         & +K_2(\overline v -v_j),
\end{array}
\label{eqn:model_equation}\\
&&\ \ \overline u \equiv {1 \over N} \sum_{i=1}^{N} u_i \mbox{\quad and\quad}
\overline v \equiv {1 \over N} \sum_{i=1}^{N} v_i.\nonumber
\end{eqnarray}
Here, two component $u_j$ and $v_j$ of $j$-th unit are considered as
activator and inhibitor by assuming that $a$,$b$,$c$, and $d$ are positive.
Each unit couples with all other units
through the averaged field $\overline u$ and $\overline v$.
We assumed that both $K_1$ and $K_2$ are non-negative.
They can be regarded as the susceptibility of each component
because this type of global coupling can be considered
as the fast limit of diffusion velocity.

First, we investigate the properties of individual unit
by setting $K_1=K_2=0$.
Steady stationary solutions $(u_0, v_0)$ are easily solved
by setting $\dot u = \dot v = 0$.
Depend on the parameter $s \equiv ad - bc$,
the number and the stability of the fixed point changed.
Linear stability of these fixed points can be analyzed by setting
$u=u_0+\delta u$,$v=v_0+\delta v$,$|\delta u|,|\delta v|\ll 1$,
and $\delta u = \delta u_0 e^{\lambda t}, \delta v = \delta v_0 e^{\lambda t}$.
Linearization of (\ref{eqn:model_equation}) leads
the eigenvalue equation
\begin{equation}
0 = \lambda ^2 -(a-d-3u_0^2) \lambda -s+3du_0^2.
\end{equation}
The fixed point $(u_0,v_0)$ is stable if the conditions
$0 > a-d-3u_0^2$ and $0 < -s+3du_0^2$ are satisfies.
We assume that $a<d$ and $s<0$ from now.
In this case, the trivial solution $(0,0)$ is the unique attractor
of the dynamical system of one individual unit.
\cite{othercase}.

Let us consider a one cluster state
which is defined as a state that every unit has the same value
$(u_{(1)},v_{(1)})$, i.e.
\begin{displaymath}
(u_j,v_j)=(u_{(1)},v_{(1)}) \mbox{\quad for \quad} j=1,...,N.
\end{displaymath}
It is generally difficult to analyze the stability of cluster state
in the globally coupled system because we must solve
eigenvalue problem of $2N$ dimensional matrix.
Following analyzing method, however, we obtain
sufficient condition for destabilization of a cluster state easily.
First, $N$ units are assumed to form $M$ clusters state
$\{(u_{(i)},v_{(i)})\},i=1,...,M$.
Next, we consider one additive unit
(test unit) $(u(t),v(t))$ in that $M$ clusters state and
make an approximation that
the effect to the test unit from $N$ units
is simply external force.
This approximation
is justified in the limit of $N\rightarrow\infty$.
By investigating the stability of this test unit,
we argued the stability condition of original $M$ clusters state as follows.
Noting that the test unit has at least
$M$ ``entrained'' solutions to each cluster,
i.e., $(u(t),v(t))=(u_{(i)},v_{(i)})$,
the linear stability of these entrained solutions can be analyzed.
If one of the entrained solutions is unstable,
we conclude that the original cluster state is unstable.
We named this stability analysis method of
the cluster state Test Unit Analysis (TUA)
\cite{TUmodeATmode}.

Now, we carry out TUA for the one cluster state $\{(0,0)\}$
given in the previous paragraph, i.e.,
we consider a stability of test unit $(u(t),v(t))$
in the external force created by the $N$ units in one cluster state.
In this case both of the average fields
$\overline u$ and $\overline v$ vanish and
equations for the test unit are written in the form:
\begin{equation}
\begin{array}{rcccl}
\dot u & = & (a-K_1)u - bv & - u^3, \\
\dot v & = & cu - (d+K_2)v & .
\end{array}
\label{eqn:test_unit_in_1c}
\end{equation}
Fixed point of test unit can be obtained by setting $\dot u = \dot v = 0$.
Using $v=cu/(d+K_2)$ given from (\ref{eqn:test_unit_in_1c}),
$u$ satisfies
\begin{displaymath}
0 = h_1(u) \equiv (d+K_2)u^3+((d+K_2)(K_1-a)+bc)u.
\end{displaymath}
Note that $h_1(0)=0$ because the test unit has the entrained solution
$(u,v)=(0,0)$.
To investigate the stability of one cluster state,
we analyze the test unit linear stability around entrained solution $(0,0)$.
By setting $u=\delta u$ and $v=\delta v$,
linear stability analysis of (\ref{eqn:test_unit_in_1c})
leads the eigenvalue equation:
\begin{equation}
0 = \lambda ^2 -(a-d-K_1-K_2) \lambda -(a-K_1)(d+K_2)+bc.
\label{eqn:test_unit_eigenvalue_in_1c}
\end{equation}
The stability conditions of the entrained solution of the test unit
to the one cluster state are now given as:
\begin{eqnarray}
0 &>& a - d - K_1- K_2\label{eqn:test_unit_stability1}\\
0 &<& -s-aK_2+K_1(d+K_2). \label{eqn:test_unit_stability2}
\end{eqnarray}
{}From the condition for existence and stability of one cluster
solution in the non-coupling case, (\ref{eqn:test_unit_stability1})
is automatically satisfied .
Therefore the critical condition for the stability is given by R.H.S of
(\ref{eqn:test_unit_stability2}) equals to $0$
where a pitch-fork bifurcation occurs.
Although this stability condition is for the
entrained solution of the test unit,
it is obvious that the original one cluster solution is unstable
if (\ref{eqn:test_unit_stability2}) is broken.
{}From the fact mentioned above, one cluster state is linearly unstable
when $K_2 > K_{2c}$.
Fig.\ 1 shows a result of numerical simulation for $K_2<K_{2c}$.
Parameters are $N=100$, $a=0.4$, $b=1$, $c=0.5$, $d=1$, $K_1=0$,
and $K_2=0.2$.
Simple Euler method with $dt=0.01$ are adopted.
One cluster state (Fig.\ 1(c))
is realized from a uniform random initial condition (Fig.\ 1(b)).
When $K_2 > K_{2c}$, the one cluster state becomes unstable and
each unit separates into two subpopulations, i.e.,
\begin{displaymath}
(u_i,v_i) = \left\{
\begin{array}{ll}
(u_{(1)},v_{(1)}) & \mbox{for }i=1,...,N_{(1)} \\
(u_{(2)},v_{(2)}) & \mbox{for }i=N_{(1)}+1,...,N
\end{array}
\right.
\end{displaymath}
Here, $N_{(1)}$ is a number of units which belongs to the first cluster.
This state is defined as two clusters state and
we focus on it in the next paragraph.

For simplicity we assumed that $K_1 = 0$.
Let a two clusters state $(u_{(1)},v_{(1)})$,$(u_{(2)},v_{(2)})$
having its proportion $p:1-p$,
here $p$ is defined as $p\equiv N_{(1)}/N$ and $0<p<1$ are satisfied.
Because the averaged fields $\overline u$ and $\overline v$ become
$pu_{(1)}+(1-p)u_{(2)}$ and $pv_{(1)}+(1-p)v_{(2)}$, respectively,
$u_{(1)},v_{(1)},u_{(2)},v_{(2)}$ satisfy the following equations:
\begin{displaymath}
\begin{array}{rcl}
0 & = & au_{(1)} - bv_{(1)} - u_{(1)}^3 \\
0 & = & cu_{(1)} - dv_{(1)} + K_2(1-p)(v_{(2)}-v_{(1)}) \\
0 & = & au_{(2)} - bv_{(2)} - u_{(2)}^3 \\
0 & = & cu_{(2)} - dv_{(2)} + K_2 p(v_{(1)}-v_{(2)})
\end{array}
\end{displaymath}
Eliminating $v_{(1)}$ and $v_{(2)}$,
transforming from $(u_{(1)},u_{(2)})$ to $(u_{(1)},\phi)$
where $\phi$ is a new variable defined by $\phi \equiv u_{(2)}/u_{(1)}$,
one can get two equations which are easily analyzed:
\begin{eqnarray*}
u_{(1)}^2 & = & {s+aK_2(1-p)-aK_2(1-p)\phi \over d+K_2(1-p)-K_2(1-p)\phi^3}\\
          & = & {(s+aK_2p)\phi-aK_2p \over (d+K_2p)\phi^3-K_2p}.
\end{eqnarray*}
$\phi$ obeys
\begin{displaymath}
(\phi-1)(bcK_2(1-p)\phi^3-s(d+K_2)\phi^2-s(d+K_2)\phi+bcK_2p)=0.
\end{displaymath}
Note that $\phi=1$ expresses $u_{(1)}=u_{(2)}$, i.e., one cluster state.
The solution must satisfies the inequality $u_{(1)}^2 > 0$.
Therefore the condition for existence of two clusters solution is :
\begin{equation}
K_2 > K_{2c} \equiv -s/a.
\label{eqn:condition_for_existence}
\end{equation}

Under the condition (\ref{eqn:condition_for_existence}),
we use TUA again to analyze the stability of the two clusters state, i.e.,
we consider a stability of test unit entrained solution
in the external force created by the $N$ units in the two clusters state.
The test unit equations are:
\begin{displaymath}
\begin{array}{rcl}
\dot u & = & au - bv - u^3 \\
\dot v & = & cu - dv + K_2(p v_{(1)} + (1-p)v_{(2)} - v).
\end{array}
\end{displaymath}
By setting $\dot u = \dot v = 0$, we obtain fixed points of the test unit.
Note that there exist two entrained solutions to the cluster
$(u_{(1)},v_{(1)})$ and $(u_{(2)},v_{(2)})$.
To investigate the stability of two cluster state,
we analyze the test unit linear stability around entrained
solution $(u_{(1)},v_{(1)})$.
By setting $u=u_{(1)}+\delta u, v=v_{(1)}+\delta v$,
linearization around $(u_{(1)}, v_{(1)})$
leads the eigenvalue equation:
\begin{equation}
0 = \lambda ^2 -(a-d-K_2-3u_{(1)}^2)\lambda-(a-3 u_{(1)}^2)(d+K_2)+bc.
\label{eqn:test_unit_eigenvalue_in_2c}
\end{equation}
The entrained solution becomes unstable if
the constant term in (\ref{eqn:test_unit_eigenvalue_in_2c})
becomes $0$.
The stability condition is
\begin{eqnarray}
0 &>& \{(aK_2+s)((2ad-3bc)K_2+2ds)^2(d+K_2p)\} \nonumber\\
  & & \times \{-9bcK_2p+2(ad+3bc)K_2+2ds\}.
\label{eqn:pK2}
\end{eqnarray}
{}From the fact that first braces of R.H.S. of (\ref{eqn:pK2})
are positive definite,
the last braces determine the stability.
Bifurcation line is given by solving about $p$, i.e.,
\begin{equation}
p = p_c(K_2) = {1 \over 9bc}({2ds \over K_2} + 2(ad+3bc)),
\label{eqn:p_critical}
\end{equation}
where a transcritical bifurcation occurs.
For example, $p_c = 7.6/9 - 0.2/4.5K_2$ for the case $a=0.4,b=1,c=0.5,d=1$.
Typical phase diagram is shown in Fig.\ 2.
In the case of $K_2 < K_{2c}$, there does not exist two clusters solution.
In the case of $K_2 > K_{2c}$, on the other hand,
one of the two clusters solution with a proportion lain
in the region B is realized.
The two clusters state in the region C
is linearly unstable and is not realized.
Note that the bifurcation diagram is symmetric to $p=0.5$ because
the proportion of the other cluster is $1-p$.
Therefore possible proportion has a minimum $p_{\rm min}= 1-p_c(K_2)$ and
maximum $p_{\rm max}= p_c(K_2)$ value for given $K_2$.
To investigate the dynamical process
we perform numerical simulations.
Fig.\ 3 shows the formation of two clusters state from
a uniform random initial condition with $K_2 = 0.3 > K_{2c}$.
Other parameters are the same as in Fig.\ 1.
At $T=500$, two clusters state is selected with a proportion $p:1-p=49:51$.
In Fig.\ 4,
the proportion regulation under artificial partial extinction is shown.
We start with the relaxed state of previous simulation (shown in Fig.\ 3(c))
with removing the 49 units which belongs to the
cluster with negative $u$ value (hatched in Fig.\ 4(a)).
The remained 51 units make a two clusters state with
proportion $p:1-p=23:28$ in the region B again.
Three or more clusters state have not been observed.

To clarify what selects the final state proportion,
we perform numerical simulations of
equations (\ref{eqn:model_equation})
with changing initial conditions.
Parameters are the same as in Fig.\ 1 except $K_2=0.5$.
We start from 500 initial conditions which are
uniform random numbers between $-0.1$ and $0.1$ with different seeds.
Fig.\ 5 shows a distribution of finally selected proportion value
with a peak at $p=0.5$.
These results show that the initial condition determines the proportion
between the two clusters.

Finally,
we check a structural stability of these result
by adding a small positive constant term $\epsilon$, i.e.,
\begin{displaymath}
\begin{array}{rccccl}
\dot u_j & = & au_j - bv_j & - u_j^3 & +K_1(\overline u -u_j) &,\\
\dot v_j & = & cu_j - dv_j &         & +K_2(\overline v -v_j) &+\epsilon .
\end{array}
\end{displaymath}
Numerical simulation shows
that the distribution of proportion also has a finite width.
The most probable proportion, however, moves from $0.5$
because the added term $\epsilon$ breaks
the symmetry of (\ref{eqn:model_equation})\cite{bifurcation}.
For example, the most probable value is about $0.25$,
$p_{\rm min}\sim0.1$, and $p_{\rm max}\sim0.4$, respectively,
when $a=0.6$, $b=1$, $c=2$, $d=1$, $K_1=0$, $K_2=3$, and $\epsilon=0.2$.
{}From this result we conclude that the proportion regulation phenomena
discussed in this letter is generic.

The DIF which regulate the proportion of two types cells are widely studied
about the differentiation process of Dd,
e.g., cyclic adenosine 3$^\prime$-monophosphate (cAMP), ammonia,
and concentration of cation are known as candidates.
If we specify the inhibitor and
if we control the susceptibility,
the proportion between two kind of cells is expected to be controlled.


The authors wish to thank N.Nakagawa, Y.Kuramoto, and Y.Sawada
for many fruitful discussions.
The present work is supported in part by the Japanese Grant-in-Aid for Science
Research Fund from the Ministry of Education, Science and Culture(No.00060267).



\begin{figure}
\caption{
(a) Temporal evolution of the distribution of units with respect to $u$.
Gray scale represents the number of units by changing
{}from $0$ to $N$ between white and black.
Randomly distributed units aggregate into the origin and
one cluster state is formed.
(b) Initial distribution of units.
Each unit has a uniform random number between $-0.1$ and $0.1$
for $u$ and $v$, respectively.
(c) Snapshot of one clusters state at $T=200$.
}
\end{figure}


\begin{figure}
\caption{
Typical phase diagram of one and two clusters state.
The upper line is given by (10) and the lower line denotes $1-p_c$.
The dotted line is $K_2=K_{2c}$.
In the region A, there is no two clusters solution.
In B and C, there is a linearly stable and unstable two clusters solution,
respectively.
}
\end{figure}


\begin{figure}
\caption{
(a) Evolution of the distribution with respect to $u$.
Units around the origin separate into two clusters state.
(b) Initial distribution.
Each unit has a uniform random number between $-0.01$ and $0.01$
for $u$ and $v$, respectively.
(c) Snapshot of two clusters state at $T=500$.
}
\end{figure}


\begin{figure}
\caption{
(a) Evolution of the distribution.
After a sudden decrease of $u$ of each unit to zero till $T\sim 10$,
51 units separate into two clusters again.
(b) Initial distribution.
49 units in the hatched cluster are removed.
(c) Snapshot at $T=500$.
Proportion $p:1-p=23:28$ is selected.
}
\end{figure}


\begin{figure}
\caption{
Probability distribution of selected proportion of two cluster state
{}from 500 randomly chosen initial conditions.
}
\end{figure}

%
%


\begin{thebibliography}{99}

\bibitem{Raper}
K.B.Raper, J.Elisha, Mitchell Scient.Soc. {\bf 56}, 241 (1940).

\bibitem{Bonner}
J.T.Bonner, Q.Rev.Biol. {\bf 32}, 232 (1957);
J.T.Bonner,
{\it The Cellular Slime Molds}, 2nd ed.Princeton Univ.Press, Princeton,
New Jersey (1967).

\bibitem{Loomis}
W.F.Loomis, {\it Dictyostelium discoideum: A Developmental System},
Academic Press, New York (1975);
K.L.Fosnaugh and W.f.Loomis, Devel.Biol. {\bf 157}, 38 (1993).

\bibitem{Takeuchi}
M.Oyama, K.Okamoto, and I.Takeuchi,
J.Embryol.exp. Morph. {\bf 75}, 293 (1983).

\bibitem{Ozaki}
T.Ozaki et al, Development, {\bf 117}, 1299 (1993).

\bibitem{Frisch}
K.von Frisch, {\it AUS DEM LEBEN DER BIENEN}, 8th edition
Springer-Verlag, Berlin and Heidelberg (1969).

\bibitem{Wilson}
E.O.Wilson, {\it The insect societies}, Oxford Univ. Press, London (1971);
G.F.Oster and E.O.Wilson, {\it Caste and ecology in the social insects},
Princeton Univ. Press, Princeton (1978).

\bibitem{Turing}
A.M.Turing, Phil.Trans.Soc. {\bf B237}, 37 (1952).

\bibitem{Kaneko}
K.Kaneko, Phys.Rev.Lett. {\bf 63}, 219 (1989); Physica D {\bf 41}, 137 (1990);
Physica D {\bf 54}, 5 (1991); Physica D {\bf 55}, 368 (1992).

\bibitem{Wiesenfeld}
K. Wiesenfeld and P. Hadley, Phys.Rev.Lett. {\bf 62}, 1335 (1989).

\bibitem{Kuramoto}
Y.Kuramoto, {\it Chemical Oscillations, Waves and Turbulence}
Springer, Berlin (1984).

\bibitem{Golomb}
D.Golomb, D.Hansel, B.Shraiman, and H.Sompolinsky,
Phys.Rev. A{\bf 45}, 3516 (1992).

\bibitem{Hakim}
V.Hakim and W.J.Rappel, Phys.Rev. A{\bf 46}, R7347 (1993).

\bibitem{Okuda}
K.Okuda Physica D {\bf 63}, 424 (1993).

\bibitem{Nakagawa_Kuramoto}
N.Nakagawa and Y.Kuramoto, Prog.Ther.Phys. {\bf 89}, 313 (1993);
Physica D {\bf 75}, 74 (1994); Physica D {\bf 80}, 307 (1995).

\bibitem{Sakaguchi}
H.Sakaguchi, to be appeared in Prog.Theor.Phys.

\bibitem{othercase}
The dynamics of each unit varies depend on the parameters.
Except $(u,v)=(0,0)$ there exists a pair of fixed points if $s \ge 0$.
Hopf and pitch-fork bifurcation occur at $a-d=0$ and $s=0$, respectively.
Around the co-dimension two bifurcation point,
a global bifurcation occurs and
coexistence of limit cycle and fixed points is observed.
More detail structure, however, will be studied in the future.

\bibitem{TUmodeATmode}
Note that the condition given by the TUA is only sufficient condition for
the destabilization of the original cluster state
and is not necessary condition.
There is an example that
the original cluster state is unstable even if TUA is stable.
Consider a following globally coupled system:
\begin{displaymath}
\dot u_j = u_j - u_j^3 + K(\overline u - u_j).
\end{displaymath}
Although the fixed point $u_j=0$ is unstable
to the perturbation that every units move to the same direction,
the entrained solution to $0$ of the test unit is stable so as to $K>0$.
The relation between the stabilities of the entrained test unit solution
and the stability of the original cluster state remains as a further problem.

\bibitem{bifurcation}
The added term $\epsilon$ also change the bifurcation type
between one and two clusters state.
Coexistence of one cluster solution and two clusters
solution occurs in some range of $K_2$.

\end{thebibliography}
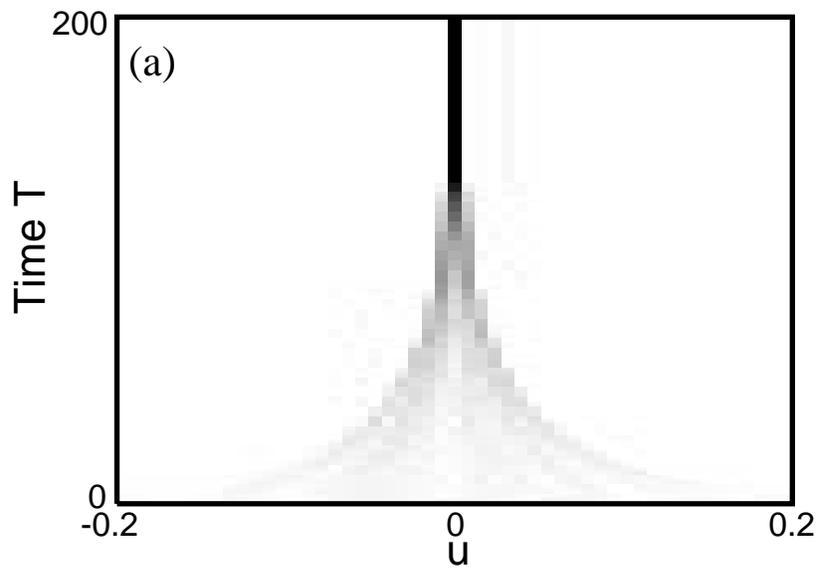
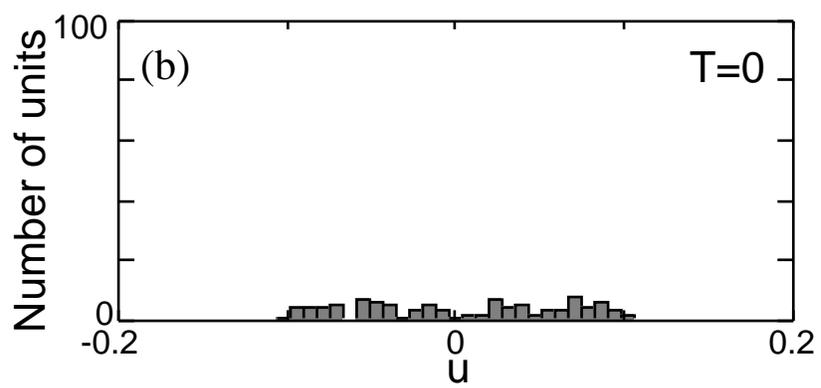
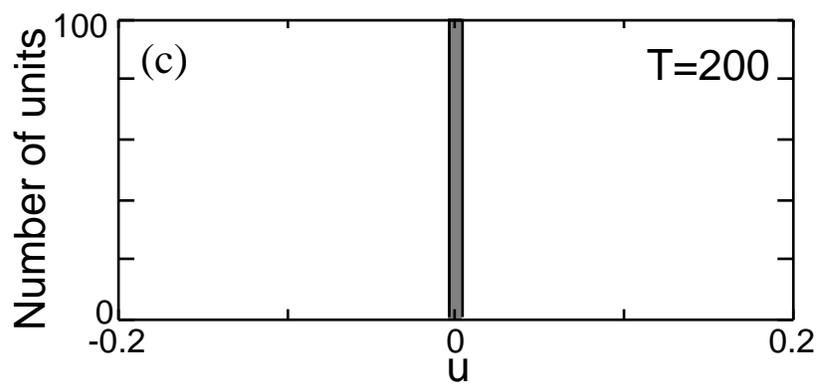
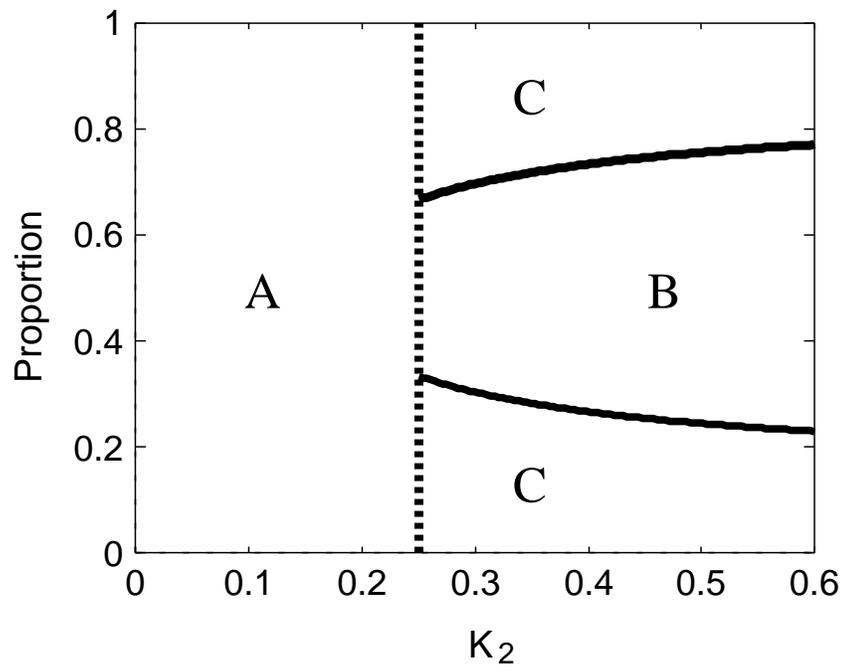
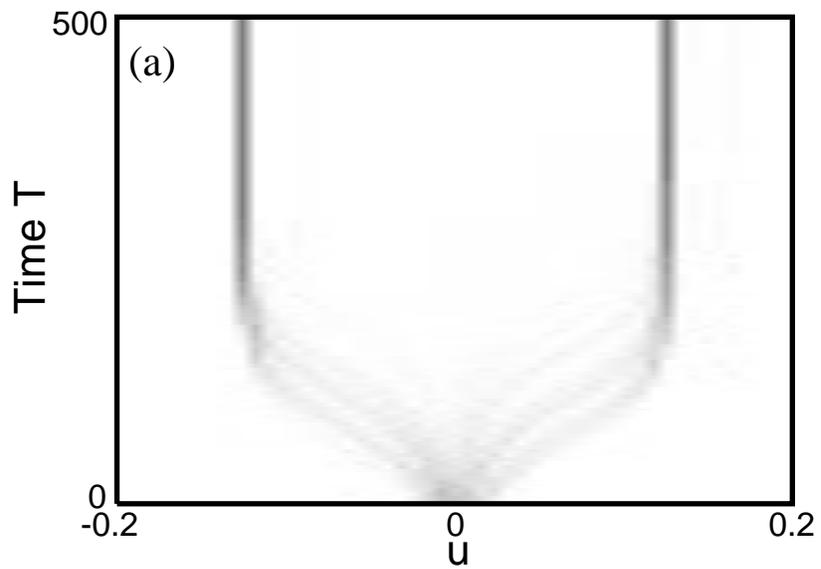
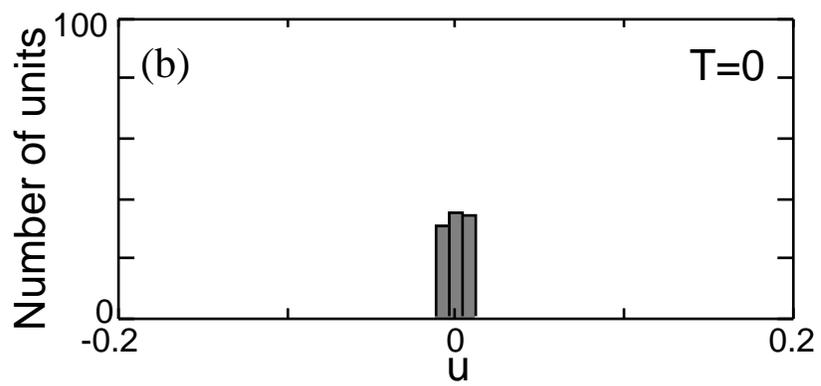
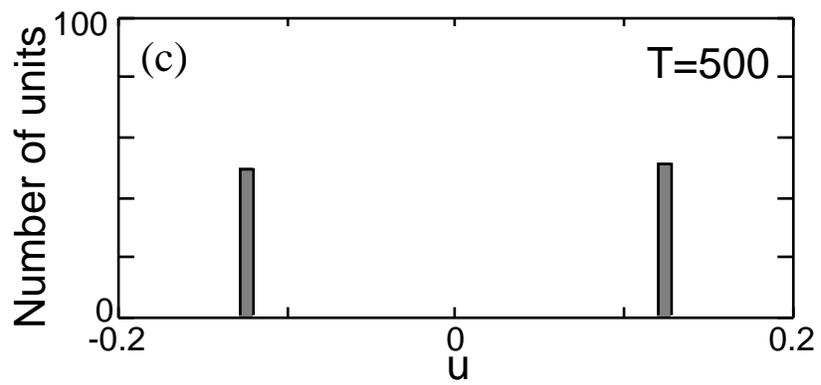
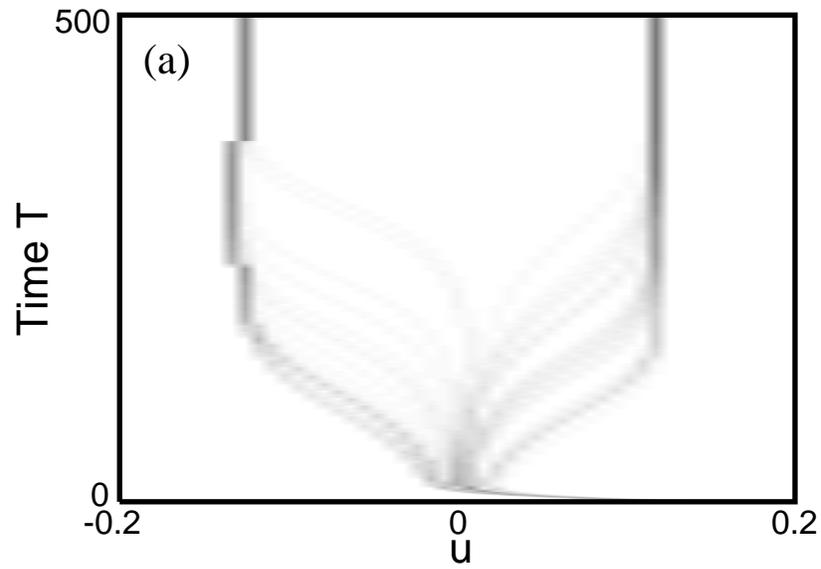
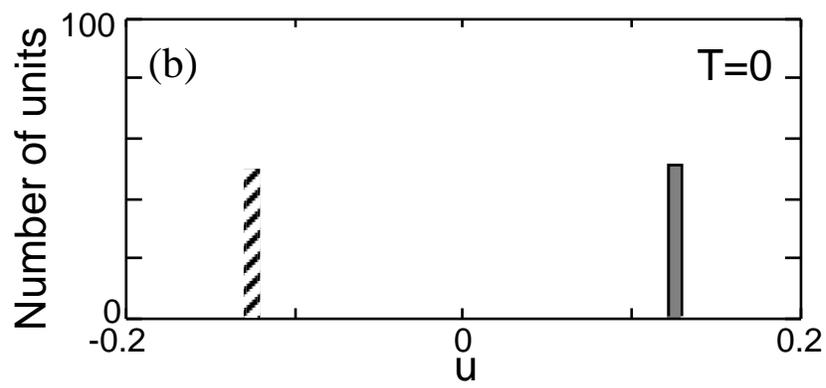
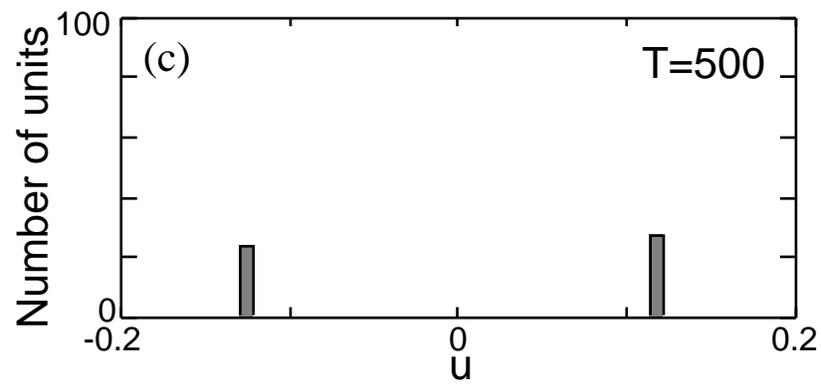
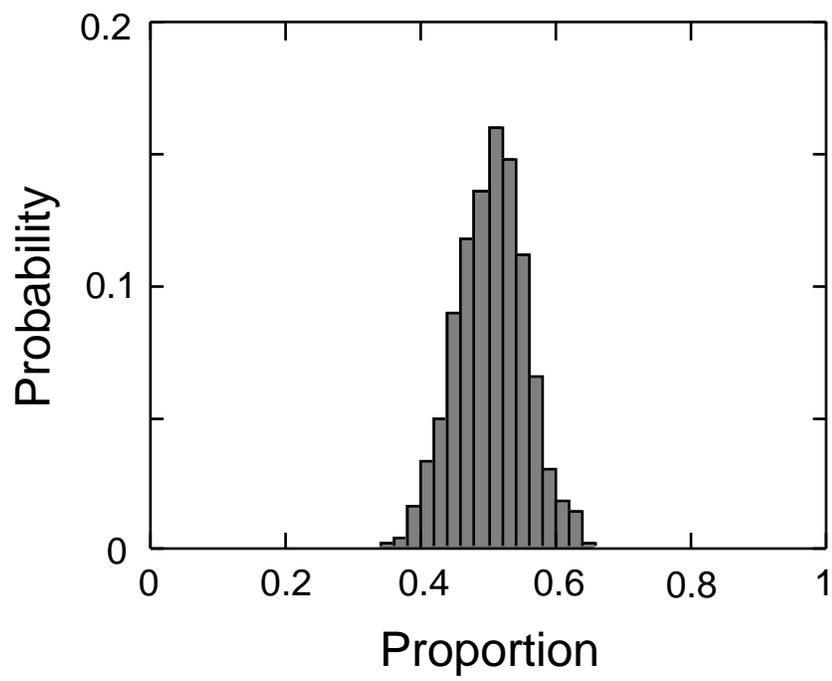
\end{document}